\documentclass{astlb}

\usepackage[utf8]{inputenc}
\usepackage[english]{babel}
\usepackage{graphicx}
\usepackage{float} 
\usepackage{wrapfig}
\usepackage[dvipsnames]{xcolor}
\usepackage{bm}
\usepackage{amsmath}
\usepackage{caption, subcaption}
\usepackage{booktabs}
\usepackage{siunitx}
\usepackage{ulem}
\usepackage{amssymb}
\usepackage{tikz}
\usepackage{mathtools, cuted}
\usepackage{calrsfs}
\usepackage{natbib}
\usepackage{xspace}
\usetikzlibrary{trees}
\usetikzlibrary{arrows.meta, chains, positioning, quotes}
\DeclareMathAlphabet{\pazocal}{OMS}{zplm}{m}{n}
\usepackage[hyperfootnotes=false]{hyperref}

\usepackage{float}
\usepackage{multicol}
\usepackage{url}

\newcommand{\srg}{{SRG}\xspace}

\newcommand{\srge}{{SRG/eROSITA}\xspace}

\newcommand{\ergcms}{erg/s/cm$^2$}

\begin{document}

\journalinfo{2022}{11}{}{1}[0]
\title{Optical identification of X-ray sources in the SRG/eROSITA Survey of Lockman Hole}

\author{
S.D. Bykov\address{1, 2} \email{sergei.d.bykov@gmail.com},
M.I. Belvedersky\address{3,4},
M.R. Gilfanov\address{3,2},
  \addresstext{1}{Kazan Federal University, Department of Astronomy and Satellite Geodesy, 420008 Kazan, Russia}
  \addresstext{2}{Max Planck Institute for Astrophysics, Karl-Schwarzschild-Str 1, Garching b. Muenchen D-85741, Germany}
  \addresstext{3}{Space Research Institute, Russian Academy of Sciences, Profsoyuznaya 84/32, 117997 Moscow, Russia}
  \addresstext{4}{National Research University Higher School of Economics (HSE), Moscow, Russia}
}

\shortauthor{Bykov, Belvedersky \& Gilfanov}
\shorttitle{Optical identification of SRG/eROSITA sources in Lockman Hole survey}
\submitted{22.11.2022}

\begin{abstract}
We present a method for the optical identification of sources detected in wide-field X-ray sky surveys. We have constructed and trained a neural network model to characterise the photometric attributes of the populations of optical counterparts of X-ray sources and optical field objects. The photometric information processing result is used for the probabilistic cross-match of X-ray sources with optical DESI Legacy Imaging Surveys sources. The efficiency of the method is illustrated using the SRG/eROSITA Survey of Lockman Hole. To estimate the accuracy of the method, we have produced a validation sample based on the Chandra and XMM-Newton catalogues of X-ray sources. The cross-match precision in our method reaches 94\% for the entire X-ray catalogue and 97\% for sources with a flux $F_{\rm x, 0.5-2}>10^{-14}$\ergcms. We discuss the further development of the optical identification model and the steps needed for its application to the SRG/eROSITA all-sky survey data.


\end{abstract}

\section{Introduction}

The SRG orbital X-ray observatory \srg \citep{sunyaev2021} was launched in July 2019 and began the all-sky survey in December 2019. There are two instruments onboard the observatory: the X-ray telescope \srge  \citep{predehl2021} sensitive in the 0.3–9.0 keV energy band and the Mikhail Pavlinsky SRG/ART-XC telescope \citep{pavlinsky2021} operating in the 6–30 keV energy band. It is expected that based on the results of its four-year sky survey, SRG/eROSITA will discover of the order of three million active galactic nuclei, about one hundred thousand clusters and groups of galaxies, and many X-ray bright stars and galaxies \citep{Prokopenko2009, Merloni2012, Pillepich2012, Kolodzig2013a, Kolodzig2013b}.

For a complete use of the X-ray sky survey results it is necessary to determine the properties of objects not only in X-rays, but also in other wavelengths. First of all, we are talking about the determination of the object type and distance. The problem is solved by identification of X-ray sources with known objects in the optical (predominantly) and other spectral ranges. This problem is known as cross-match \citep{Sutherland1992, Budavavari2015, Pineau2017}. In this paper we consider the problem of searching for optical counterparts of X-ray sources. We will call the X-ray sources “sources” and the candidates for optical counterparts “objects.” A peculiarity of the problem is that the positional error of X-ray sources is larger than the positional error of optical objects approximately by an order of magnitude. In addition, because of the high density of optical objects, there may be several candidates for optical counterparts in in X-ray source positional uncertainty region. In this case, it is often difficult to determine which of the optical candidates is the true counterpart (or that there is no counterpart at all). Additional data, for example, the magnitudes or colours of objects, need to be used to solve this problem. This paper is devoted to solving this problem using the SRG/eROSITA data as an example.

One of the earliest approaches to the inclusion of additional information was demonstrated in \citet{Sutherland1992} using the likelihood ratio method (LR). The method takes into account the positions of sources in the catalogues (including the errors), their angular number density, and the magnitude distribution. Subsequently, this method was developed further in \citet{Budavari2008, Naylor2013, Budavavari2015,  Pineau2017, Salvato2018} with a generalisation to the cross-correlation of more than two catalogues and the construction of a probabilistic (Bayesian) model to take into account the set of measurements of any type.

The history of applying various methods of searching for counterparts to X-ray surveys (predominantly the Chandra and XMM-Newton surveys) is extensive \citep{brusa2007, Luo2010, Naylor2013,  Marchesi2016,  Ananna2017, Luo2017, Pineau2017,  Salvato2018, Chen2018, Belvedersky2022, Salvato2022}. In the broad sense different methods use similar approaches to the problem. In the first step the distributions of magnitudes (and, sometimes, also colours or other attributes) are calculated for the optical counterparts and field objects, and in the second step this information is combined with the information about the relative positions of the sources from different catalogues and their positional errors. The achieved cross- match accuracy depends on the specific method of including the photometric information and the quality of the X-ray and optical catalogues, their depth and positional uncertainties.

The aim of this paper is to solve the problem of the optical cross-match of X-ray sources for the parameters of the SRG/eROSITA all-sky survey \citep[see also][]{Belvedersky2022}. In this paper we describe the cross-match method and its application to the deep Lockman Hole survey data and estimate the accuracy of the proposed method using a validation sample. Our main goal is to present and validate the method of cross-matching SRG/eROSITA X-ray sources. The characteristics of the X-ray observations, the catalogue of sources, and the results of their optical cross-match and classification are presented in the accompanying papers of the series of publications devoted to the deep SRG/eROSITA Lockman Hole survey (Gilfanov et al. 2023, in press; Belvedersky et al. 2022, in press).

 The paper is structured as follows. In Section \ref{sect-data} we describe the data used. In Section \ref{sect-model} we construct the cross-match model and produce the training and validation samples of objects. In Section \ref{sect-results}  we discuss the results and conclude.

\section{Data}\label{sect-data}

\subsection{X-ray data}\label{sect-data-xray}

The Lockman Hole is the area  of low interstellar absorption \citep{lockman1986} on the sky that was studied by eROSITA within the performance verification phase of SRG telescopes in the fall of 2019. A relatively deep survey of this region was carried out in November 2019. The survey footprint is  $5\degr~\times~3.7\degr$ in size with an area $18.5$ square degree (for the flux limit of $\sim 5\times10^{-15}$ \ergcms\,). The exposure depth was 8 ks per point; the sensitivity in the 0.3–2.3 keV energy band is $\sim~3\times10^{-15}$ \ergcms\,. The centre of the region has approximate coordinates $\text{RA}~=~162 \degr; \text{Dec}~=~58\degr$. Further details of the observations, the X-ray data, and the catalogue of sources are given in Gilfanov et al. (2023).

The optical cross-match model was developed and tested on the catalogue of medium- and high- brightness eROSITA Lockman Hole X-ray sources (Gilfanov et al. 2023). Point sources with a detection likelihood $DL>10$ were included in this catalogue ($DL>10$ roughly corresponds to significance $\approx 4\sigma$,). The catalogue includes 6885 X-ray sources.

For the construction and validation of the optical identification model we used data from the Chandra and XMM-Newton X-ray observatories. We took the Chandra data from the Chandra Source Catalogue 2.0, CSC \citep{Evans2020}, and the XMM-Newton data from the 4XMM DR10 catalogue \citep{webb2020}. These catalogues were filtered by detection quality flags to minimise the number of spurious X-ray sources\footnote{Filters for CSC: conf\_flag, extent\_flag, sat\_src\_flag, pileup\_flag, dither\_warning\_flag are all false, $\text{likelihood}>10$. 4XMM DR10 filtering: sc\_sum\_flag$=$0 or 1, also sc\_var\_flag, sc\_extent and confused equal to false, sc\_det\_ml$>10$.}. In the SRG/eROSITA Lockman Hole survey footprint we found 2029 and 1316 Chandra and XMM sources, respectively. The data of these observatories cover about 20\% of the footprint area. The positions of the SRG/eROSITA, Chandra, and XMM-Newton sources in the Lockman Hole field are shown in Gilfanov et al. (2023).

\subsection{Optical data}
\label{sect-data-optical}

We chose the photometric DESI Legacy Imaging Survey \citep[DESI LIS, ][]{dey2019} as a catalogue of optical sources among which we will search for counterparts to SRG/eROSITA sources. DESI LIS has sufficient sensitivity and large area\footnote{\url{https://www.legacysurvey.org}}. The catalogue is produced from the data of three telescopes: BASS (g, r), MzLS (z), and DECaLS (g, r, z). These three telescopes have different limiting sensitivities in different filters. DECaLS provides data at $\text{Dec}<32\degr$. Hence, for the Lockman Hole ($\text{Dec}\approx58\degr$) and our study of the photometric attributes we used the BASS and MzLS data. In addition, the DESI LIS catalogue provides data on the infrared fluxes in four WISE filters. All magnitudes were dereddened.

The DESI data were used in the entire Lockman Hole field (with a mean density of objects $\sim~58000$ sources per degree squared ) to identify the eROSITA sources. For training the photometric model, we used all DESI objects around the CSC 2.0 sources in the sky with $\text{Dec}>32\degr$ (see Subsection \ref{sect-model-phot-train}).

For all DESI objects we calculated the signal-to-noise ratio (S/N) for the measurements in all filters (g,r,z,w1,w2,w3,w4). If the measurement in some filter had S/N less than 3, then we assumed that the measurement in this filter is absent.

\section{Model}\label{sect-model}

In this section we describe the photometric model for the selection of candidates for optical identification and its application to the eROSITA data. We describe the photometric model in Subsection \ref{sect-model-phot} or, more specifically, the training sample in Subsection \ref{sect-model-phot-train} and the photometric neural network classifier in Subsection \ref{sect-model-phot-magnn}. We discuss the validation sample in the Lockman Hole field in Subsection \ref{sect-model-validation} and the application of the Bayesian approach to the cross-match problem and its results for the eROSITA sources in Subsection \ref{sect-model-nway}.

\subsection{Photometric model}\label{sect-model-phot}

In this part we study the photometric attributes of two populations: 
\begin{enumerate}
\item 
the DESI LIS objects that are the true counterparts of X-ray sources (in our case, Chandra sources)
\item 
the DESI LIS objects that are not the counterparts of X-ray sources, the so-called field optical sources. 
\end{enumerate}
This division of the optical objects into two classes is motivated by the fact that the X-ray sources are a fairly rare and peculiar type of objects, so that their optical counterparts have attributes different from those of field objects.

For the selection of optical objects into the first class we used unambiguous optical counterparts of X-ray sources from the Chandra catalogue (which has high positional accuracy). The procedure for the selection of sources with unambiguous optical counterparts will be described below.

\subsubsection{Training sample}
\label{sect-model-phot-train}

We used the CSC 2.0 data in the extragalactic sky ($|b|>20\degr$) within the DESI LIS coverage zone, excluding the Lockman Hole area. The produced sample of 157958 Chandra sources was cross-correlated with the DESI catalogue with a search radius of 30 arcsec. Thus, we selected $\sim3$ million DESI objects. For each Chandra source we determined the local density of optical objects $\rho_{\rm desi}$ based on their number in a 10–30 arcsec ring. To control the frequency of random chance associations between CSC and DESI, for each X-ray source we calculated the chance identification radius $r_{\rm false}$ from the formula
$$r_{\rm false} = \sqrt{-\frac{{\rm ln}(1-f_{\rm lim})}{\pi \rho_{\rm desi}} }$$
where $f_{\rm lim}$ is the probability to find one or more optical objects at a distance $r<r_{\rm false}$ from the X-ray source \citep{Belvedersky2022}. The quantity $r_{\rm false}$ was calculated by assuming a homogeneous distribution of optical objects with a density $\rho_{\rm desi}$. As a threshold probability we chose $f_{\rm lim} = 0.03$. For each source from the Chandra catalogue we also calculate the positional uncertainty $r_{98}$ corresponding to the radius within which the true X-ray source position is located with a 98\% probability.

Our further selection of reliable optical counterparts was made as follows: if only one DESI LIS object was found within $r_{\rm false}$ from the X-ray source, with the separation between the two being smaller than $r_{98}$, then this object was deemed \textit{a reliable optical counterpart of the Chandra source}. We also left only  reliable Chandra sources (see Subsection \ref{sect-data-xray}). We found 71 993 such Chandra– DESI pairs.

For the selection of \textit{reliable field sources} we found all DESI objects in a 10–30 arcsec ring from the Chandra source. The number of field objects was 2 135 168.

In the next step, in the sample of reliable counterparts we left only those objects for which the $0.5-2$ keV X-ray flux from the Chandra source was larger than $3\times 10^{-15}$ \ergcms, corresponding to the approximate eROSITA limiting flux in the Lockman Hole field. Note that this is a very important step, since the training and validation samples must be equivalent in their properties to the X-ray catalogue for which the optical identification will be made. For example, if we included all Chandra sources in the sample of reliable counterparts, then their photometric attributes would be biased toward numerous faint sources inaccessible to SRG/eROSITA \citep{Salvato2018}.

Finally, we selected only reliable counterparts and field objects located at $Dec>32\degr$ (see Subsection \ref{sect-data-optical}). From the field sources obtained in this way we randomly selected only half, since the total number of objects turned out to be large.

As a result of this selection procedure, the size of the sample of reliable counterparts was 12 452, while the size of the sample of reliable field objects was 244 008.

\begin{figure}
    \centering
    \includegraphics[width=\columnwidth]{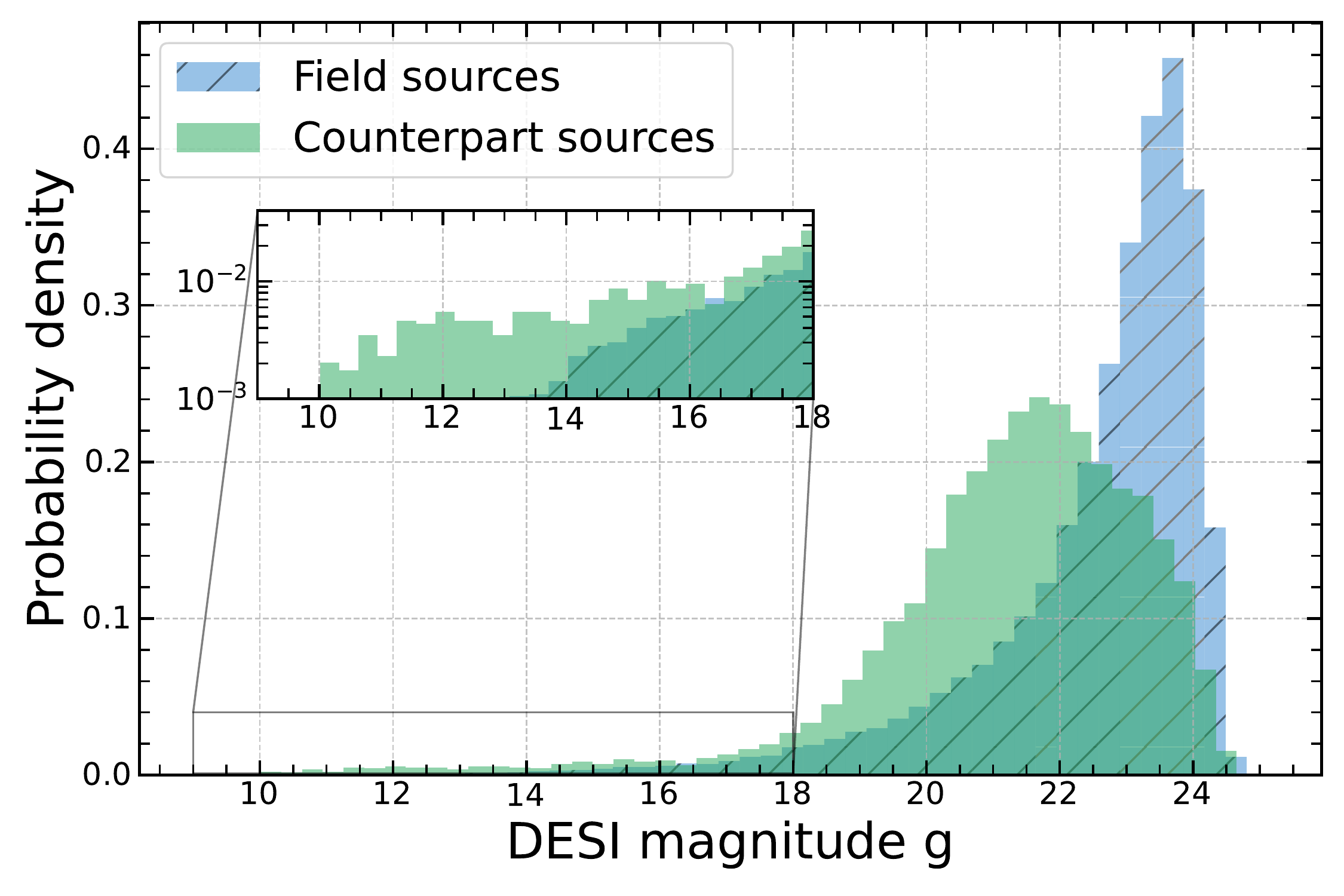}
    \caption{The distribution of apparent g magnitude  for the DESI objects that fell into the samples of counterparts (green histogram) and field sources (blue histogram with dashes). The inset shows the region of bright objects on a logarithmic scale along the y axis.
    }
    \label{fig:mag_g}
\end{figure}

In Fig. \ref{fig:mag_g} we show the distribution of the sources from these two samples in DESI LIS g filter apparent magnitude. It can be seen that, on average, the counterparts of the X-ray sources are brighter than the field sources. In the absence of additional information, this fact can be used to determine more preferable optical counterparts to eROSITA sources (for a discussion, see below).

The limiting S/N in the training sample (see Subsection \ref{sect-data-optical}) was changed from 3 to 4.

\subsubsection{Photometric classifier}\label{sect-model-phot-magnn}

The problem of classification is to determine whether a source is more likely the counterpart of an X-ray source or more likely a field source based on a set of photometric information. Such problems are efficiently solved with machine learning algorithms.

As attributes for the model we used the g, r, z, w1, and w2 magnitudes and the appropriate colours. We did not use the magnitudes w3 and w4, since these measurements are lacking for 95 and 90\% of the sources in the Lockman Hole, respectively. In addition, for some of the X-ray sources there are no measurements in the g, r, z, w1, and w2 filters (18, 12, 6, 35, and 60\%, respectively). For this reason, we introduced three models: the first model uses only the attributes (magnitudes and colours) pertaining to the g, r, and z filters, the second model uses the g, r, z, and w1 filters, and the third model uses g, r, z, w1, and w2.

As a classifier we used a neural network, with the corresponding attributes as input and  a number between 0 and 1 that we called nnmag as output; 1 and 0 were assigned to counterparts and field objects, respectively. Below we will describe the model based on all attributes: g, r, z, w1, and w2.

We constructed a neural network with $\operatorname{tanh}$  activation function in which there were 4 layers of 8 neurons each. After each layer we added a dropout layer with a probability of 0.1 to prevent over-fitting. At the output of the neural network we added a layer with one neuron with logistic activation. As a loss function we used binary cross-entropy. The number of sources of the “counterpart” and “field” classes was 10 480 and 55 636, respectively; we used 30\% of the data as test sample, 20\% for cross-validation, and the remainder as a training sample. \textsc{Keras}\footnote{\url{https://keras.io}} package was used to train the classifiers.

\begin{figure}
    \centering

    \includegraphics[width=1\columnwidth]{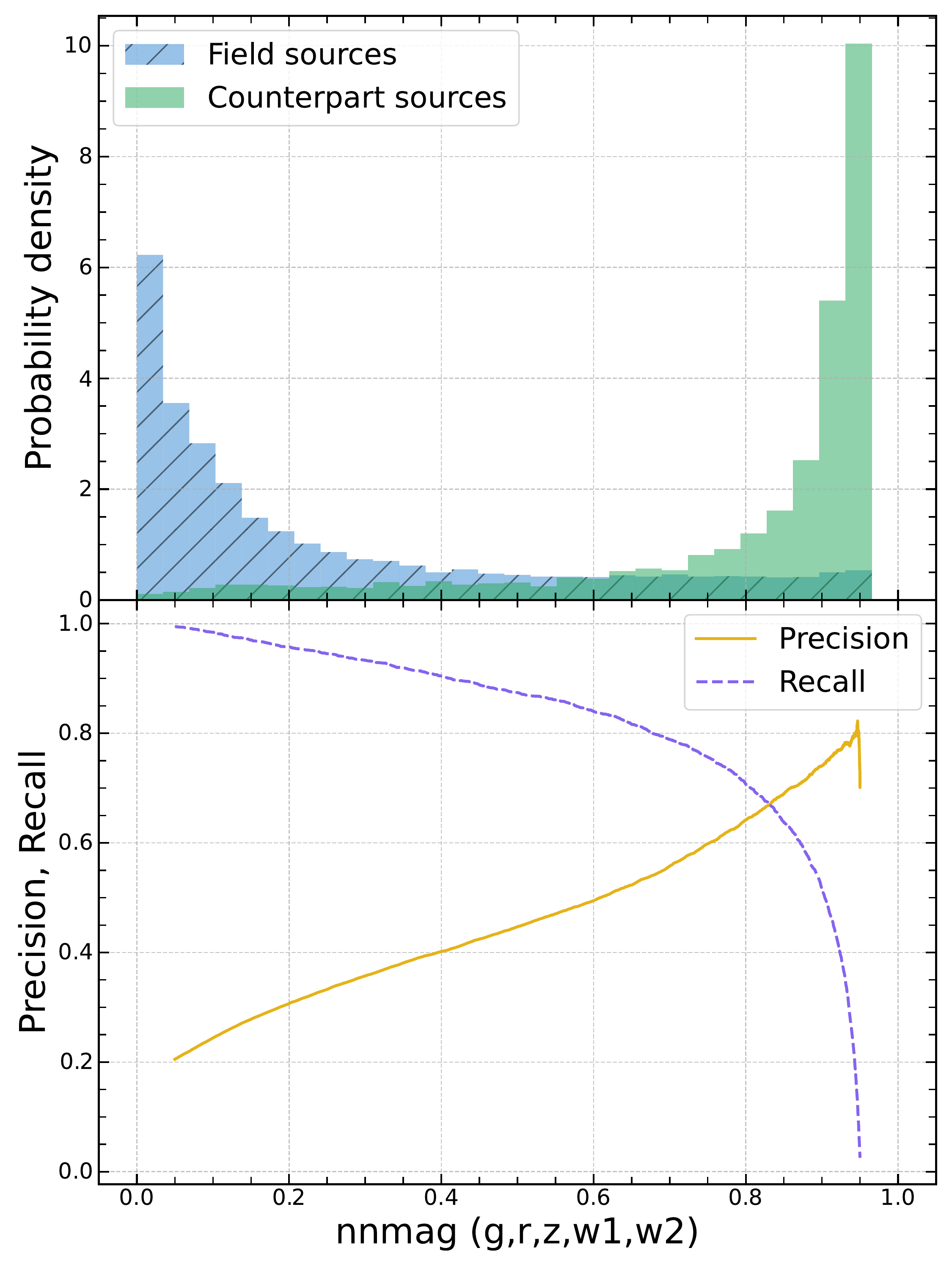}
    \caption{Upper panel: the distribution of neural network output (nnmag) for the DESI objects that fell into in the test samples of field sources (blue histogram with dashes) and counterparts (green histogram). Lower panel: the recall (dashed line) and precision (solid line) curves for the counterparts as a function of nnmag threshold.}
    \label{fig:nnmag_grzw1w2}
\end{figure}

Figure \ref{fig:nnmag_grzw1w2} (upper panel) illustrates a histogram of the neural network output (nnmag) on the test sample (the data that were not used for training) for both classes of objects. It can be seen that the objects of class 1 (counterparts) have nnmag closer to 1, while the objects of class 0 (field sources) have nnmag closer to 0. This histogram can be compared with Fig. \ref{fig:mag_g}, where the separation in g magnitude is not so obvious. The lower panel in Fig. \ref{fig:nnmag_grzw1w2} also shows two model quality metrics for different nnmag thresholds: recall and precision. At a given nnmag the sources are designated as a counterpart (field) if their nnmag is larger (smaller) than the threshold \footnote{The recall is calculated from the formula $TP/(TP+FN)$, while the precision is calculated as $TP/(TP+FP)$, where TP (true positive) is the number of correctly assigned counterparts, FN (false negative) is the number of counterparts assigned as field ones, and FP (false positive) is the number of field sources assigned as counterparts}. As an overall model quality metric we used the precision at the point of intersection of the two curves; for the g,r,z,w1,w2 model this number is 0.67. Note that at the point of intersection of the two curves the recall and the precision are equal.

Important parameters for all three models are described in Table 1: the attributes used as input features,  sample sizes, neural network structure, and the overall model quality. The drop in the model precision with the loss of data from the infrared is worth noting— the grzw1w2, grzw1, and grz models have an overall precision of 0.67, 0.61, and only 0.38, respectively.

\begin{table*}[]\label{tab:nnmags}
\begin{tabular}{c|cccc}
Model         & Features                                                                                & Sample size$^1$                                               & Network structure                                                                    & Overall quality$^2$ \\\hline\hline
nnmag\_grzw1w2 & \begin{tabular}[c]{@{}c@{}}g,r,z, w1, w2\\ g-r,r-z,g-z\\ z-w1, r-w1, w1-w2\end{tabular} & \begin{tabular}[c]{@{}c@{}}1: 10480\\ 0: 55636\end{tabular}  & \begin{tabular}[c]{@{}c@{}}4 layers with 8 neurons\\ Dropout rate: 0.1\end{tabular}  & 0.67           \\\hline
nnmag\_grzw1   & \begin{tabular}[c]{@{}c@{}}g,r,z, w1\\ g-r,r-z,g-z, z-w1\end{tabular}                     & \begin{tabular}[c]{@{}c@{}}1: 10814\\ 0: 89670\end{tabular}  & \begin{tabular}[c]{@{}c@{}}4 layers with 8 neurons\\ Dropout rate: 0.1\end{tabular}  & 0.61           \\\hline
nnmag\_grz     & \begin{tabular}[c]{@{}c@{}}g,r,z\\ g-r,r-z,g-z\end{tabular}                             & \begin{tabular}[c]{@{}c@{}}1: 11092\\ 0: 141845\end{tabular} & \begin{tabular}[c]{@{}c@{}}2 layers with 13 neurons\\ Dropout rate: 0.0\end{tabular} & 0.38      
\end{tabular}
\caption{Description of the photometric models. For each model we show the used photometric features, sample sizes, neural network structure and the overall quality of a model.
1: Class 1 and class 0 refer to the counterparts and field sources, respectively.
2: The recall at the point of intersection of the recall and precision curves (see Subsection \ref{sect-model-phot-magnn}).
}
\end{table*}

\subsection{Validation sample in the Lockman Hole}\label{sect-model-validation}

To estimate the optical identification quality, we need to have a sample of eROSITA sources with known optical counterparts. For this we invoke the Chandra and XMM-Newton data in the Lockman Hole. We used an approach similar to that in \citet{Belvedersky2022}.

The selection procedure is as follows:

\begin{enumerate}
    \item We cross-correlated the complete Chandra (XMM) catalogues \footnote{Without a quality filter, see Subsection \ref{sect-data-xray}} with the eROSITA catalogue with a search radius of 30 arcsec and wrote out the unique eROSITA–Chandra (XMM) pairs, leaving only the reliable Chandra/XMM detections from them. 461 (646) such pairs were obtained.

    \item  For each Chandra (XMM) source we found all DESI LIS objects within 15 arcsec and chose those from them that (i) lay within $1.1\times r_{98}$ from the X-ray source and are the only optical object within this radius and (ii) lay within $r_{\rm false}$. We found 310 (383) Chandra (XMM)–DESI pair (for the definition of the quantities, see Subsection \ref{sect-model-phot-train}). As $r_{\rm false}$ we chose a radius of 1.47"\ from the average DESI sky density in the Lockman Hole.

    \item Thereafter, from the Chandra (XMM)–DESI pairs we excluded those in which the 0.5– 2 keV fluxes from the eROSITA and Chandra/XMM data differed by more than a factor of 5 \citep{Belvedersky2022}— 235 (374) sources remained.

    \item  The two catalogues (eROSITA–Chandra–DESI and eROSITA–XMM–DESI) were combined, while the duplicates were removed. We obtained a catalogue consisting of 548 eROSITA–Chandra (XMM)–DESI triples. From it we removed the cases where there were one reliable XMM/Chandra source and two or more Chandra/XMM sources that did not pass the detection quality filter. 529 eROSITA–Chandra (XMM)– DESI triples remained.

\end{enumerate}

This allowed to determine the correspondence between an eROSITA X-ray source and a DESI optical object. This catalogue will be used below to estimate the optical identification quality.

We also produced a catalogue of eROSITA sources that have no optical counterpart in the DESI LIS catalogue—these are the so-called hostless sources. In the second step we changed the search criterion by requiring that there was no DESI optical object within $2\times r_{98}$ and $2\times r_{\rm false}$. The remaining steps were not changes. The final number of hostless eROSITA sources in the validation sample was 30.

\subsection{NWAY and the Identification of X-ray Sources in the Lockman Hole}\label{sect-model-nway}


The information about astrometry and photometric properties was combined with the NWAY code \footnote{\url{https://github.com/JohannesBuchner/nway}} \citep{Salvato2018}, which uses the Bayesian approach to calculate the probabilities of different candidates to be a true counterpart. Two catalogues, eROSITA and DESI, are fed to the input of this code. Initially, the match probability for each possible pair is calculated based only on the astrometric positions, their errors, and the density of sources in both catalogs. Subsequently, the probabilities are refined using additional information (hereafter, photometric priors). In this case, the refinement is based on the ratio of the prior probability densities for the counterparts and field sources. At the output with each eROSITA object we associate the number $p_{\rm any}$ characterising the presence of an optical counterpart and with each candidate DESI source  we associate the number $p_{\rm i}$ characterising its probability to be the counterpart. The closer the numbers to 1, the more reliable the corresponding assertion.

As a photometric prior we used nnmag obtained by the neural network—photometric classifier (Sub- section \ref{sect-model-phot-magnn}, see also the example in Fig. \ref{fig:nnmag_grzw1w2}). The photometric model (Table 1) was chosen, depending on which attributes were available for a particular DESI LIS object. At the same time, we made sure that there is no “leak” of information: for each optical object we used only one photometric prior from the model with the greatest precision available for this object (Table 1). If, however, the photometric information about the object was not sufficient for any of the models in Table 1 (for example, there was no measurement in the z filter), then as a prior we used any available information (magnitudes and colours) calibrated based on the sample from Subsection \ref{sect-model-phot-train}, for example, as illustrated in Fig. \ref{fig:mag_g}.

As $1\sigma$\footnote{Corresponds to a probability of 39\% for a bivariate normal distribution.} positional errors we used the corresponding calibrated positional error from the eROSITA catalogue of X-ray sources (Gilfanov et al. 2023) and a fixed error of 0.1"\ for the DESI LIS sources.

With NWAY we calculated the parameter $p_{\rm any}$ for each eROSITA source and $p_{\rm i}$ for each optical object within 30 arcsec from the X-ray source. The candidate with the greatest $p_{\rm i}$ was designated as the most probable counterpart. The X-ray sources with $p_{\rm any}$ below the threshold $p_{\rm any, 0}=0.12$ were deemed hostless. The choice of the threshold $p_{\rm any, 0}$ is discussed in the next section.

\section{Results and discussion}\label{sect-results}

In this section we present the results of the optical identification of eROSITA X-ray sources in the Lockman Hole. We provide the overall statistics on the number of identifications in Subsection \ref{sect-results-numbers}, describe the calibration of the positional errors of the X-ray catalogue based on the validation sample in Subsection \ref{sect-results-poserr}, and describe the estimation of the precision (purity), recall (completeness) of the catalogue of counterparts in Subsection \ref{sect-results-identification}. In Subsection \ref{sect-results-next} we discuss a generalisation of the model to the entire sky and the changes in the procedure needed for this.

\subsection{Identification results}
\label{sect-results-numbers}

As a result of applying the cross-match procedure described above, 6346 (92\%) and 4360 (63\%) of the 6885 eROSITA sources in the Lockman Hole field have $p_{\rm any}~>0.12$ and $p_{\rm any}~>0.8$, respectively. For 5866 and 1019 X-ray sources the nearest and not nearest DESI LIS object, respectively, was assigned a counterpart. For 6458 and 427 X-ray sources the most probable optical counterpart lies at angular separations $\operatorname{sep}<r_{98}$ and $\operatorname{sep}>r_{98}$, respectively. For 443 (6\%) eROSITA sources we found two or more candidates with similar values of $p_{\rm i}$ (within a factor of 2). For 5919 (86\%) eROSITA sources one of the optical objects has $p_{\rm i}>0.8$.

\subsection{Positional Errors of eROSITA X-ray Sources}\label{sect-results-poserr}

The 529 eROSITA–DESI LIS validation pairs allow us to check the validity of the positional error for the X-ray sources. If the errors were calibrated correctly, then the distribution of $\frac{\operatorname{sep}}{\sigma}$ (where $\operatorname{sep}$ is the separation between the observed position of the X-ray source and its true position, and $\sigma$ is the X-ray positional error) is expected to follow the Rayleigh distribution \citep{Watson2009, Chen2018, Brunner2021, Belvedersky2022}.

\begin{figure}
    \centering
    \includegraphics[width=1\columnwidth]{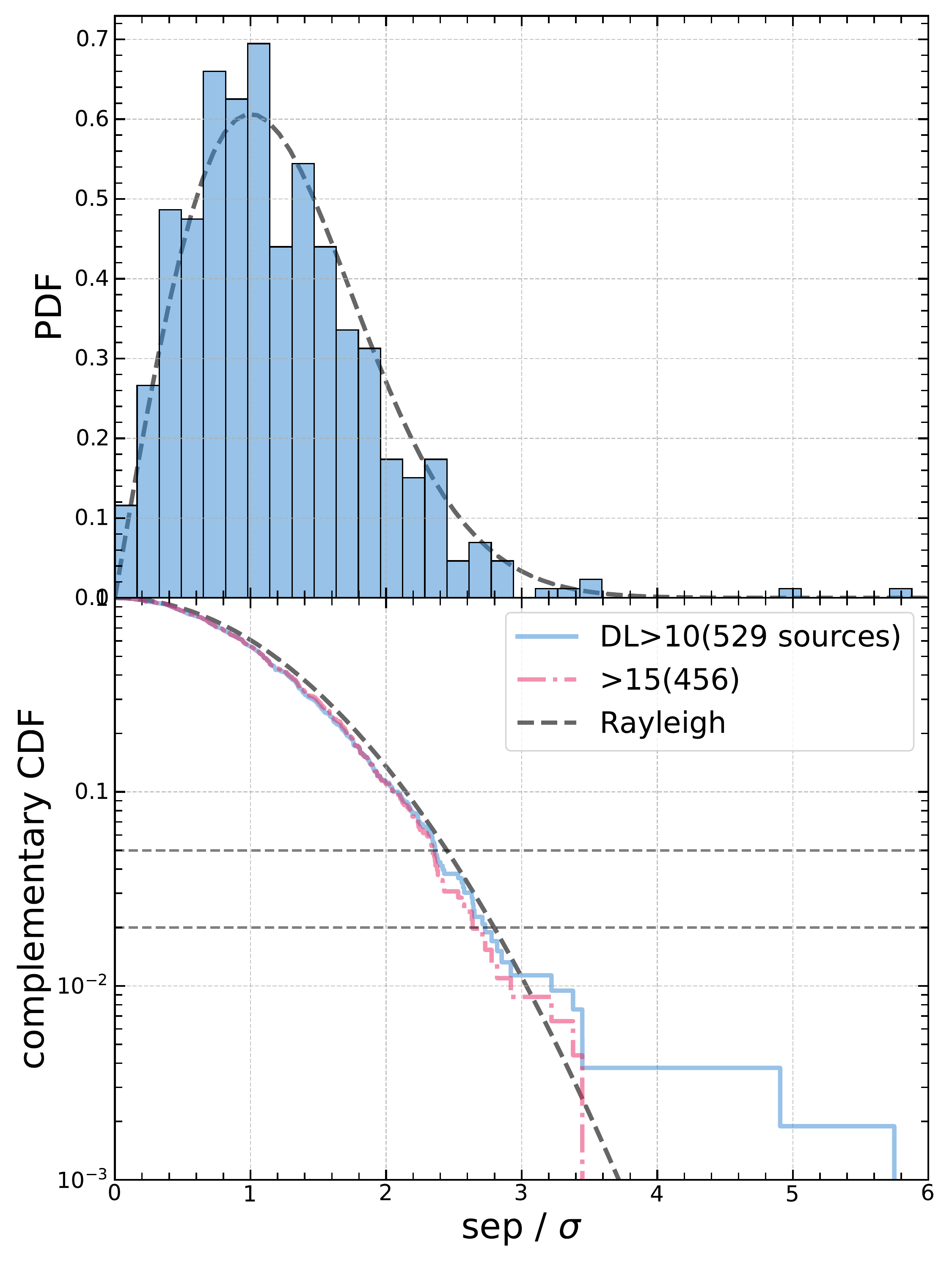}
    \caption{Distributions in offsets of the X-ray sources from their optical counterpart position in  validation sample. Upper panel:  blue histogram shows the  probability distribution of $\frac{\operatorname{sep}}{\sigma}$. Lower panel:  cumulative distribution of the same quantity for two selections of X-ray sources by detection likelihood $DL$ ($DL>10$— blue solid line, $DL>15$— red dash–dotted line). On both panels the black dashed line indicates the expected Rayleigh distribution. On the lower panel the horizontal lines indicate  95 and 98\% probability levels.}
    \label{fig:pos_err}
\end{figure}

The results are presented in Fig. \ref{fig:pos_err}, where the distribution of this quantity  from the validation catalogue is shown. As the true position of an X-ray source we used the position of its optical counterpart that is known with an accuracy better than fractions of arcsec, whilst the typical positional accuracy of X-ray sources is $\sim 3-20$ arcsec . Figure \ref{fig:pos_err}   presents the distributions for the complete catalogue in which the detection likelihood $DL>10$ and for the sources with  $DL>15$ (corresponding to a significance level $\approx 5\sigma$). The higher the threshold, the fewer the number of spurious sources. The expected Rayleigh distribution is also shown.

As can be seen from Fig. \ref{fig:pos_err}, the observed distributions follow well the Rayleigh distribution. For the complete validation sample including the sources up to $DL=10$ we can see two sources at large deviations $\operatorname{sep}\ga 5\sigma$. A visual inspection of these sources revealed no obvious anomalies in the X-ray or optical data. The appearance of these two sources is the result of a chance superposition of unrelated eROSITA and Chandra sources; some of the eROSITA sources may be false (i.e., the result of statistical fluctuations in the X-ray image), as may be suggested by their comparatively low statistical significance ($<~5\sigma$). Note that at the $4\sigma$ significance threshold we expect $\sim10-20$ false sources in the Lockman Hole (for a more detailed discussion, see Gilfanov et al. 2023). There is no such tail in the distribution of high-significance ($DL>15$) sources—it agrees well with the Rayleigh distribution.

Thus, the positional errors given in the catalogue of eROSITA sources adequately describe the uncertainties in the positions of the X-ray sources, justifying their use in calculating the Bayesian source cross- match probabilities in the NWAY code.

\subsection{Optical Identification Precision and Recall}\label{sect-results-identification}

It is necessary to estimate the quality of our identification model, to compare it with the algorithm for a trivial choice of the nearest optical object as a counterpart, and to calibrate the $p_{\rm any}$ threshold.  We use the validation sample and compare the counterparts selected by our model and the true counterparts found using the Chandra/XMM data.

To interpret the results of our cross-match model, we apply the following algorithm. Using the preselected threshold $p_{\rm any, 0}$, we classify the eROSITA sources with $p_{\rm any}~<p_{\rm any, 0}$ as hostless. For the sources with $p_{\rm any}~>p_{\rm any,0}$ we choose the object with the maximum value of $p_{\rm i}$ as a counterpart. In this case, the following outcomes are possible:

\begin{itemize}
  \item It was correctly determined that the source has a counterpart and the counterpart was chosen correctly. We denote this number by A.
  \item It was correctly determined that the source has a counterpart, but the counterpart itself was chosen incorrectly (B). This is the identification error.
  \item The hostless source was incorrectly classified as a source with a counterpart (C). This is the classification error of the hostless source.
  \item The source with a counterpart was incorrectly classified as a hostless one (D). This is the classification error of the source with a counterpart. 
  \item It was correctly determined that the source is hostless (E).
\end{itemize}

Several examples of sources from the validation catalogue and their optical surroundings are given in the Appendix in Fig. \ref{fig:match_images}. The positions of X-ray sources (eROSITA, Chandra, XMM) and  optical objects are marked. We also specify the identification parameters ($p_{\rm any}$, $p_{\rm i}$) and the true counterpart. We show three examples of correct matches and one example of incorrect match.

In the validation sample there are $N_{c}=529$ sources with counterparts and $N_{h}=30$ hostless sources.
We specify the following quality metrics \citep{Belvedersky2022}:
\begin{itemize}
  \item overall metric $Q ~= \frac{A+E}{N_{c}+N_{h}}$
 
  \item counterpart identification recall (completeness) $C_{c}~=\frac{A}{N_{c}}$
  \item hostless identification recall $C_{h}~=\frac{E}{N_{h}}$ 
  \item counterpart identification precision (purity) $P_c ~= \frac{A}{A+B+C}$ 
  \item hostless identification precision $P_h ~= \frac{E}{D+E}$ 
\end{itemize}

\begin{figure}
    \centering
    \includegraphics[width=1\columnwidth]{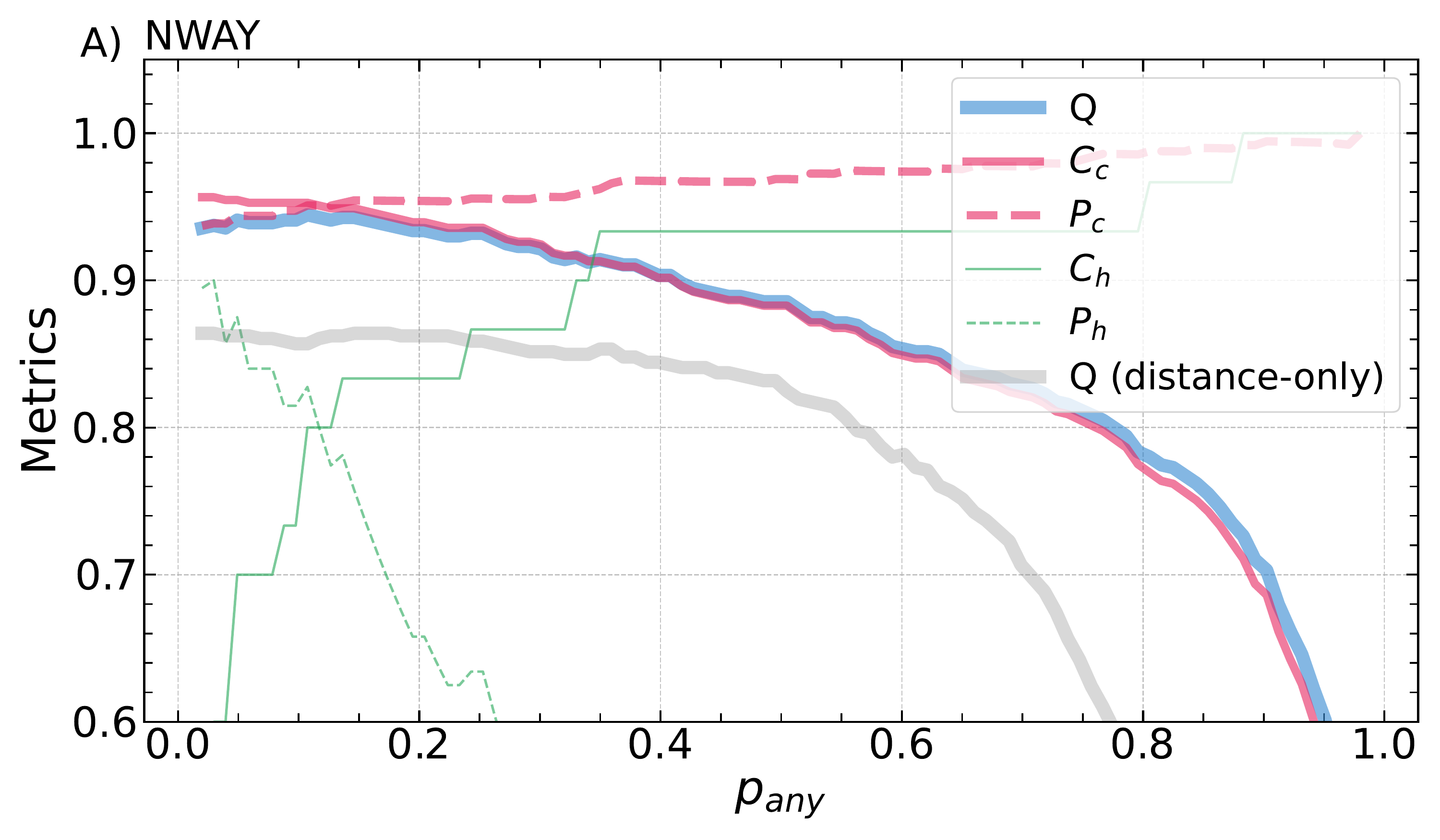}
    \includegraphics[width=1\columnwidth]{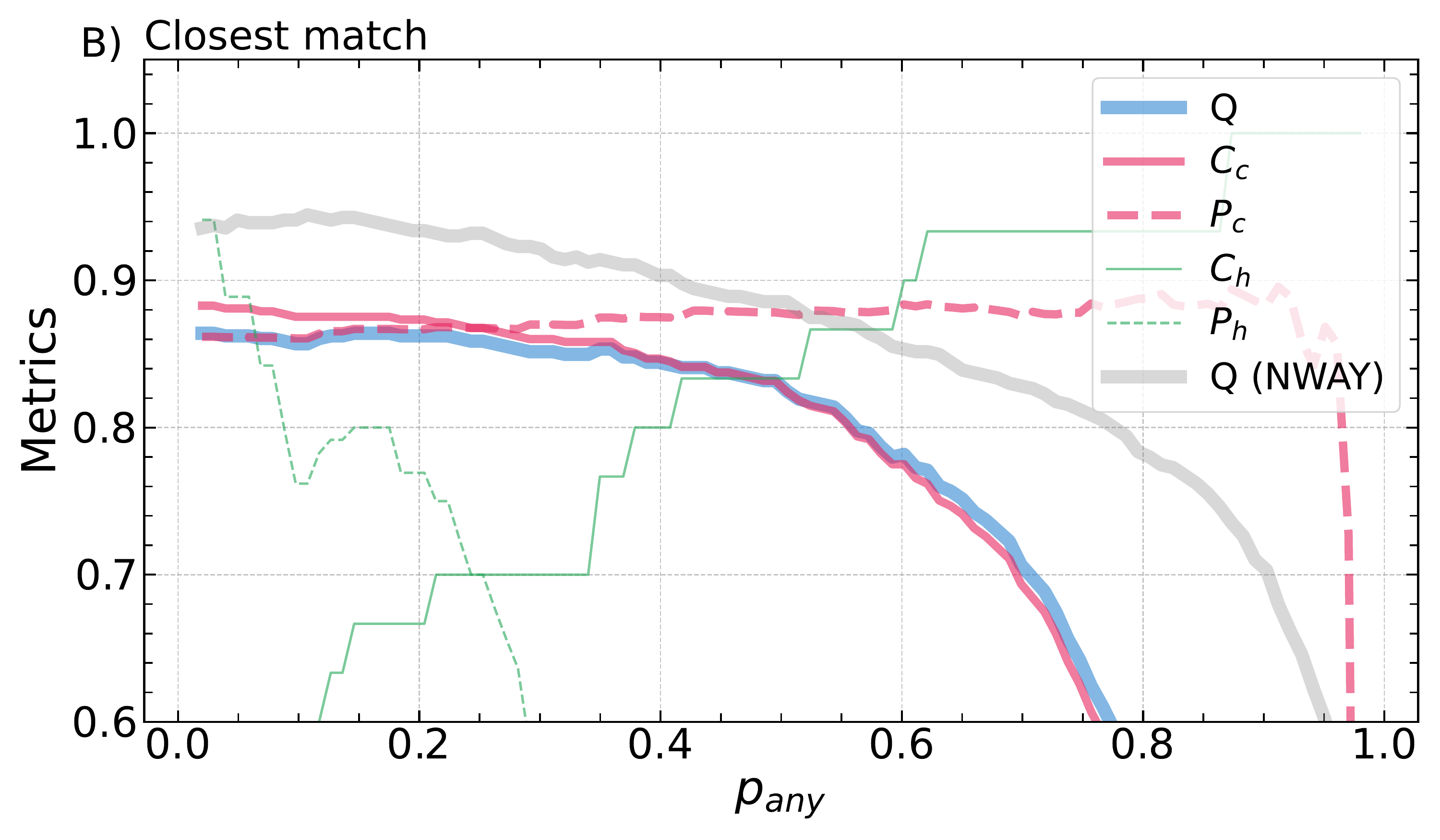}
    \includegraphics[width=1\columnwidth]{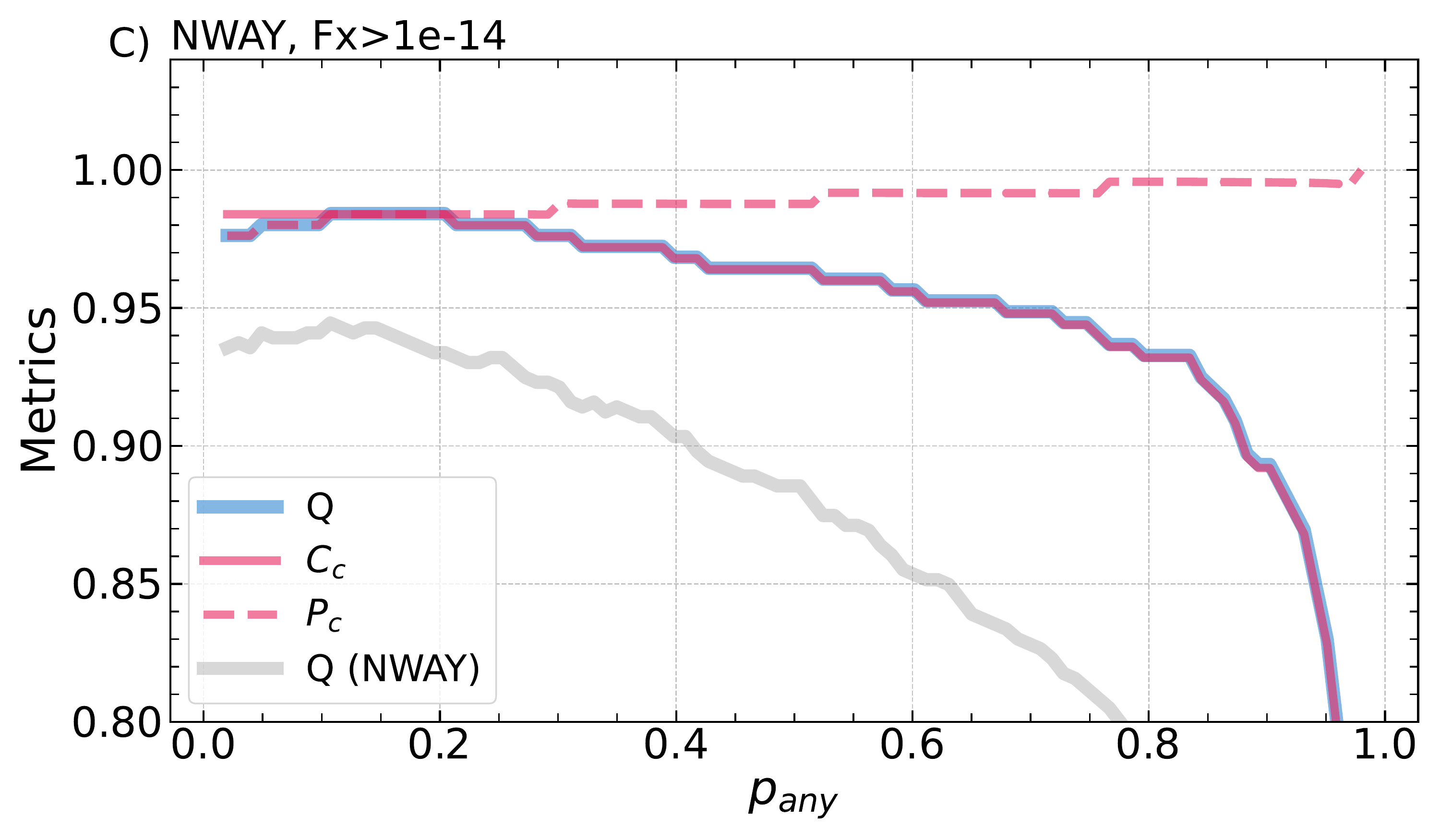}
    \caption{ Cross-match model quality metrics versus threshold $p_{\rm any, 0}$. Solid thick blue line indicates the behaviour of metric $Q$. Medium-thickness red lines correspond to the selection of counterparts: solid line is the recall $C_c$, dashed line is the precision $P_c$. Thin green lines indicate that for the selection of hostless sources: solid line is recall $P_h$, dashed line is precision $P_h$. (A) complete identification model using the photometric and positional information; grey line indicates metric $Q$ for the selection of the nearest counterpart. (B) identification with the nearest optical object; grey line indicates metric $Q$ of photometric model. (C) complete cross-match model for bright sources ($F>10^{-14}$ \ergcms); grey line indicates metric $Q$ of the photometric model for the complete catalogue.}
    \label{fig:quality_matrics}
\end{figure}

Figure \ref{fig:quality_matrics} shows the behaviour of the metrics for three versions of the identification model: the complete model using the photometric and positional information, the positional model (the nearest optical object is chosen as a counterpart), and the complete model applied to bright sources (with a 0.5-2 keV energy range flux  $F_X>10^{-14}$\ergcms).

The point of intersection of the counterpart selection recall and precision curves is located at $p_{\rm any}~=0.12$, with a precision and recall reaching 94–95\%. For the hostless sources  intersection occurs at $p_{\rm any}~\approx0.15$ with a precision and recall of 80\%. At $p_{\rm any}~=0.12$ the overall precision of our identification model reaches Q = 94\%. $p_{\rm any}$ threshold must be chosen, depending on the specific scientific problem for which the optical cross-match is made. For the general X-ray source identification/classification problem threshold $p_{\rm any, 0}~=0.12$ is a good choice.

In the case of a naive identification of an X-ray source with the nearest optical object from the DESI LIS catalogue, the counterpart selection curves intersect at $p_{\rm any}~\approx0.24$ with a precision and recall of 86– 87\%, and $p_{\rm any}$ is the same with a precision and recall of 70\% for the selection of hostless ones. The overall precision of the model is Q = 86\%.

When bright sources (0.5–2 keV flux $F_x~>~10^{-14}~$ \ergcms) are identified, the precision and recall reach $\sim98\%$ for $p_{\rm any, 0}~\approx0.12$. Note that with this X-ray flux filter there are only three hostless sources in the validation sample. Therefore, metrics for hostless sources were not calibrated.

Table 2 presents the results of the quality estimation in the corresponding cases.

\begin{table*}[]\label{tab:quality}
\small

\begin{tabular}{c|cccccc}
Model                                                                                          & \begin{tabular}[c]{@{}c@{}}Companion\\ is correct\\ (A)\end{tabular} & \begin{tabular}[c]{@{}c@{}}Companion\\ is incorrect\\ (B)\end{tabular} & \begin{tabular}[c]{@{}c@{}}Companion \\ assigned\\ to hostless\\ (C)\end{tabular} & \begin{tabular}[c]{@{}c@{}}Source \\ with companion\\ assigned hostless\\ (D)\end{tabular} & \begin{tabular}[c]{@{}c@{}}Correct\\ hostless\\ (E)\end{tabular} & \begin{tabular}[c]{@{}c@{}}Model \\ quality\end{tabular} \\ \hline
\begin{tabular}[c]{@{}c@{}}NWAY + \\ nnmag\\ $p_{\rm any, 0} = 0.12$\end{tabular}                   & 503                                                              & 20                                                                 & 6                                                                                    & 6                                                                                               & 24                                                              & Q = 0.94                                                   \\ \hline
\begin{tabular}[c]{@{}c@{}}closest\\ candidate\\ $p_{\rm any, 0} = 0.24$\end{tabular}               & 460                                                              & 61                                                                 & 9                                                                                   & 8                                                                                               & 21                                                              & Q = 0.86                                                   \\ \hline
\begin{tabular}[c]{@{}c@{}}NWAY + \\ nnmag\\ (bright sources)\\ $p_{\rm any, 0} = 0.11$\end{tabular} & 246                                                              & 4                                                                  & 0                                                                                    & 0                                                                                               & 3                                                               & Q = 0.98                                                   \\ \hline
\end{tabular}

\caption{The number of correctly or incorrectly determined counterparts of eROSITA sources for three versions of the cross-match model: the complete model, the selection of the nearest optical source, and the complete model applied to bright sources from the validation catalogue. Columns show the number of correct or incorrect identifications (see sect. \ref{sect-results-identification}), and the overall quality.}
\end{table*}

There is an increase in the identification quality when adding the photometric information— the overall cross-match quality Q increases by 8\%. Interestingly, however, the identification with the closest optical source provides a good accuracy per se. This is due to the relatively good eROSITA positional accuracy and the density of optical objects at the DESI LIS sensitivity level in the Lockman Hole. The density of DESI objects in the Lockman Hole corresponds to roughly one object in a 10"\, circle which is approximately the positional accuracy of SRG/eROSITA. For this reason, the identification of eROSITA sources with the nearest optical object in many cases leads to the correct result.

\subsection{Further Steps for the All-Sky Survey}\label{sect-results-next}

Lockman Hole Field was observed by eROSITA during the performance verification period of SRG observations. The X-ray exposure is distributed fairly uniformly over the survey area. In addition, the survey area is not too large, which allows us to ignore the interstellar absorption and DESI LIS depth variations. When the photometric model is extended to larger sky areas, it will be necessary to take into account these effects. DESI LIS does not cover the entire sky, and Pan-STARRS \footnote{\url{https://outerspace.stsci.edu/display/PANSTARRS/}} or similar optical large area surveys can probably be used as a photometric catalogue.

In Subsection \ref{sect-model-phot} we described a fairly flexible photometric model training procedure. First, in the training data we placed emphasis on the X-ray survey depth $3\times 10^{-15}$  \ergcms, and the update of the procedure to the all-sky survey regime (where the X-ray depth varies) is fairly easy and consists of partitioning the models into X-ray flux bins. Second, emphasis in this paper was put on extragalactic fields ($|b|>20\degr$), and using a new training sample from the Galactic disk region will allow one to approach the problem of the optical identification of X-ray sources in densely populated fields in the ridge of our Galaxy ($|b|>20\degr$).

A validation catalogue from the all-sky data will allow to calibrate more accurately the positional errors of X-ray sources  (see Subsection \ref{sect-results-poserr}) through a noticeable increase in the sample size. It will probably also be possible to take into account the subtler effects associated with the dependence of the positional error on the X-ray flux, the detection likelihood \citep{Belvedersky2022}, the position in the sky, etc. A large validation catalogue will help to estimate more accurately the recall and purity curves of the cross-match algorithm and its dependence on the detection parameters, including the position in the sky.

\section{CONCLUSIONS}

In this paper we constructed a model for the optical identification of sources discovered by eROSITA using the eROSITA catalogue of X-ray sources in the Lockman Hole as an example and estimated its precision and recall. As a source of optical information we used DESI LIS data. The models were trained and validated using Chandra and XMM X-ray data.

We presented a photometric model based on neural networks to separate the photometric attributes of the optical counterparts of X-ray sources and field sources. The model transforms the entire set of available photometric data into one number that we called nnmag and allows these two classes to be separated with a precision/recall of 40–70\%, depending on the available data.

We describe the construction of a sample of eROSITA sources with reliably determined optical counterparts. Such a sample allows the precision and recall of any cross-match model to be characterised and, if necessary, is easily constructed in larger sky fields. In the Lockman Hole field this allowed to pinpoint the optical counterpart for 559 of the 6885 eROSITA. The validation sample allowed us to check the accuracy of the positional errors for X-ray sources, as described in Subsection \ref{sect-results-poserr}.

We applied a combination of the neural network classifier and the NWAY code to search for optical counterparts of SRG/eROSITA sources in the Lockman Hole.

The quality of the constructed optical identification model is discussed in Subsection 4.3. The  model reaches a recall (completeness) and precision (purity) of 95\% for the selection of counterparts and 80\% for the selection of hostless sources, which exceeds the naive identification with the nearest optical source by 8\%. For bright X-ray sources ($F_{0.5-2}>10^{-14}$ \ergcms) the model reaches 98\% putiry/completeness. We give examples of identifications and detailed calibration curves. In conclusion, we discussed the necessary steps to apply the model to the SRG/eROSITA all-sky survey.

\section*{ACKNOWLEDGMENTS}

This study is based on observations with the eROSITA telescope onboard the SRG observa- tory. The SRG observatory was built by Roskosmos in the interests of the Russian Academy of Sciences represented by the Space Research Institute (IKI) within the framework of the Russian Federal Space Program, with the participation of the Deutsches Zentrum fuer Luft- und Raumfahrt (DLR). The SRG/eROSITA X-ray telescope was built by a consortium of German institutes led by MPE, and supported by DLR. The SRG spacecraft was designed, built, launched and is operated by the Lavochkin Association and its subcontractors. The science data are downlinked via the Deep Space Network Antennae in Bear Lakes, Ussurijsk, and Baykonur, funded by Roskosmos. The eROSITA data used in this paper were processed with the eSASS software developed by the German eROSITA consortium and the proprietary data reduction and analysis software developed by the Russian eROSITA Consortium.
S.D. Bykov thanks the International Max Planck Research Schools (IMPRS) at the Ludwig Maximilian University of Munich for its support. M.I. Belvedersky and M.R. Gilfanov thank the Russian Science Foundation (project no. 21-12-00343) for its support. 

The following software was used: NumPy \citep{Harris2020}, Matplotlib \citep{Hunter2007},  SciPy \citep{2020SciPy-NMeth}, Pandas\citep{reback2020pandas},  AstroPy \citep{astropy:2018}, HEALPix \citep{Gorski2005}, HEALPy 9). The code used to produce the results of the paper would be available shortly after the publication of the X-ray catalog and the corresponding paper \footnote{\url{https://github.com/SergeiDBykov/lockman_hole}}.

\vfill
\pagebreak   
 
\bibliographystyle{astl}
\bibliography{refs_rus.bib}

\section*{Appendix}

  \begin{figure*}[!ht]
\centering
    \subfloat[]{\includegraphics[width=0.45\linewidth]{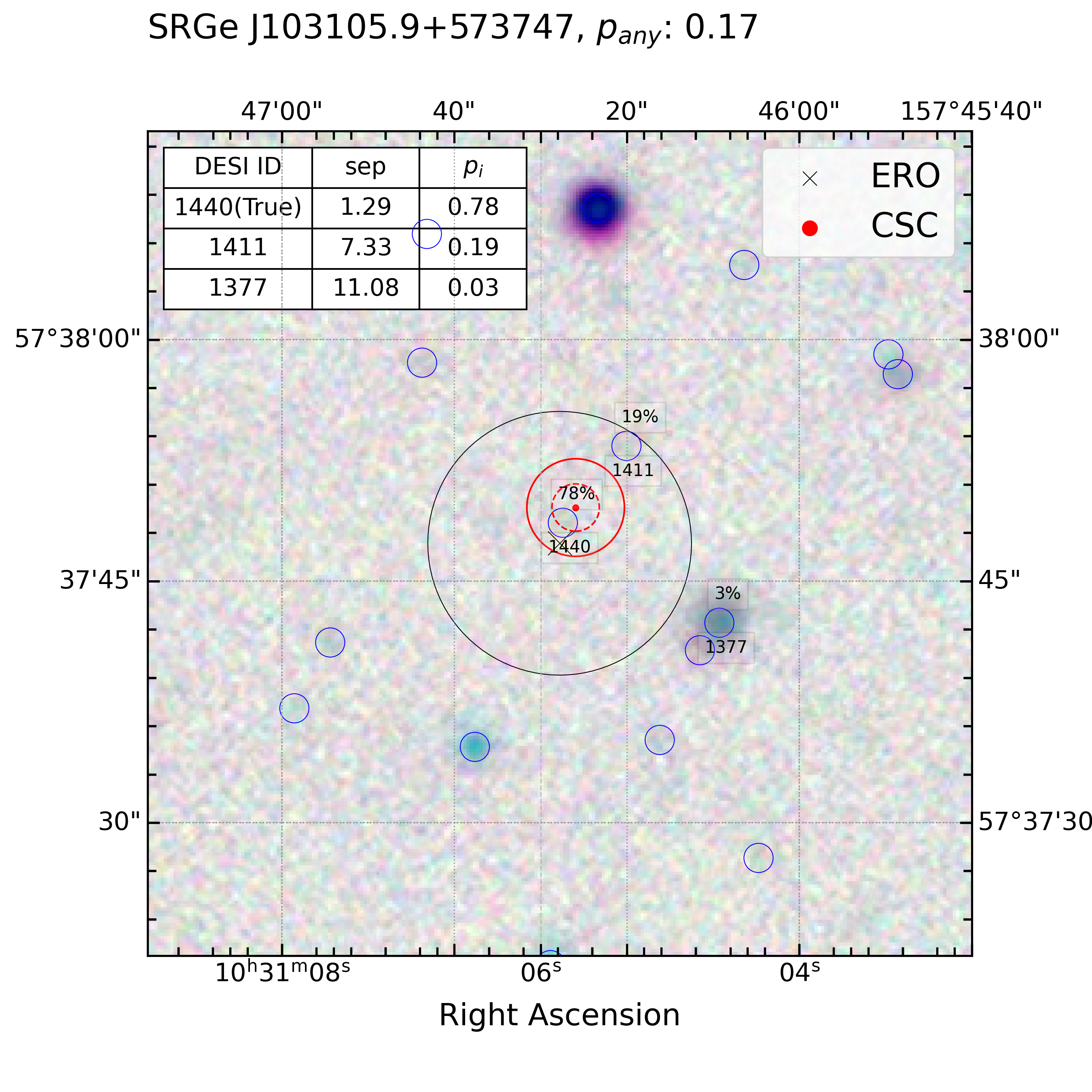}}
\hfil
    \subfloat[]{\includegraphics[width=0.45\linewidth]{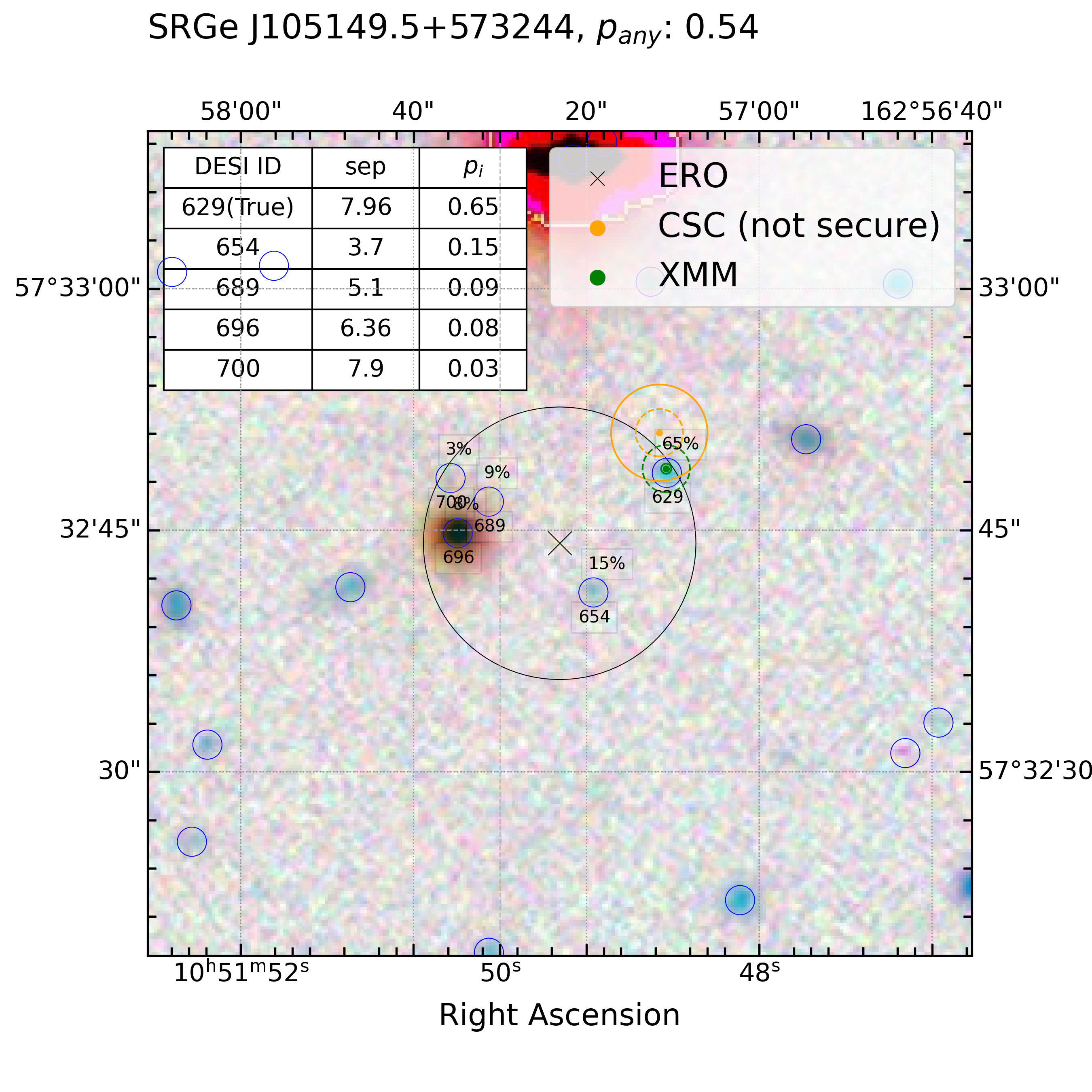}}

    \subfloat[]{\includegraphics[width=0.45\linewidth]{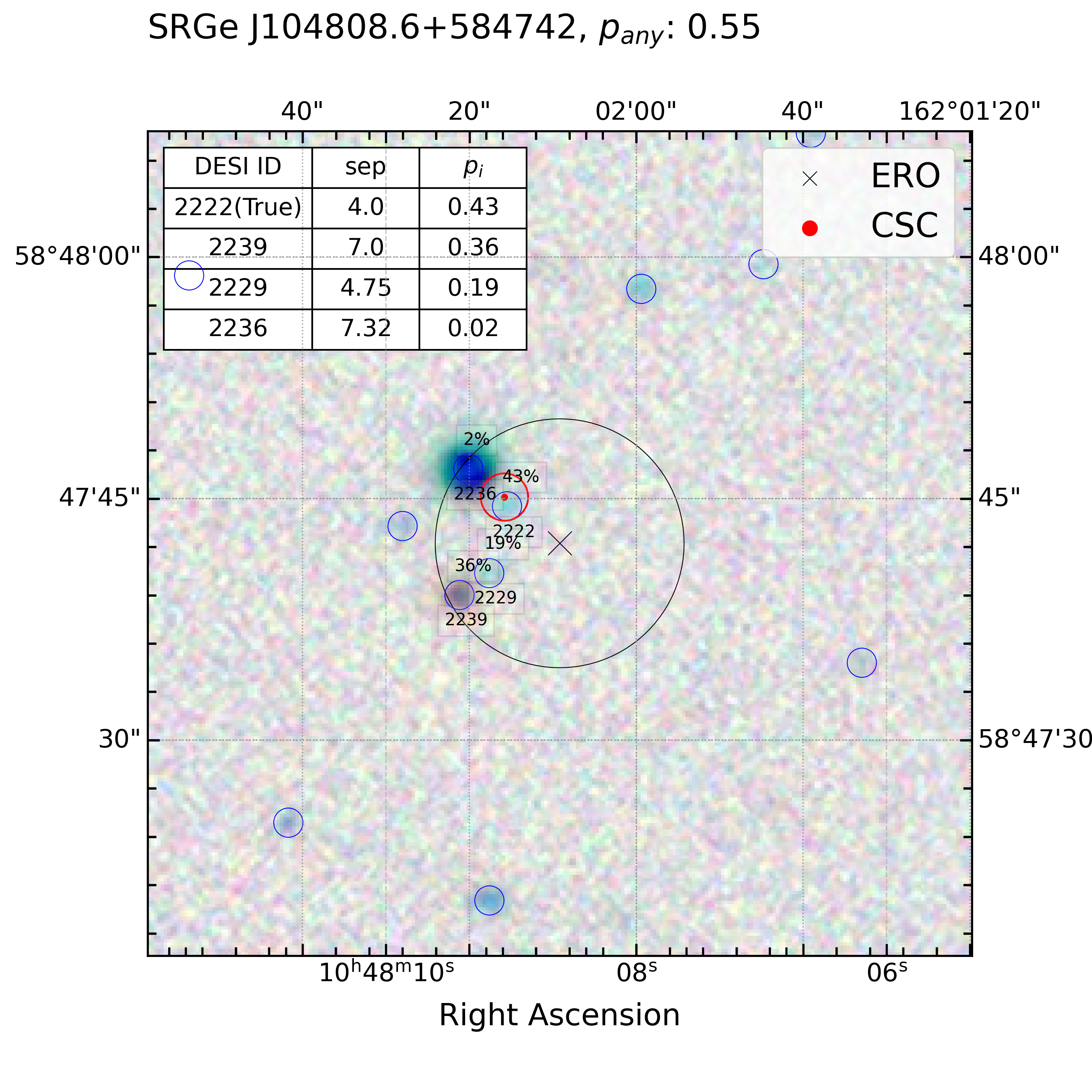}}
\hfil
    \subfloat[]{\includegraphics[width=0.45\linewidth]{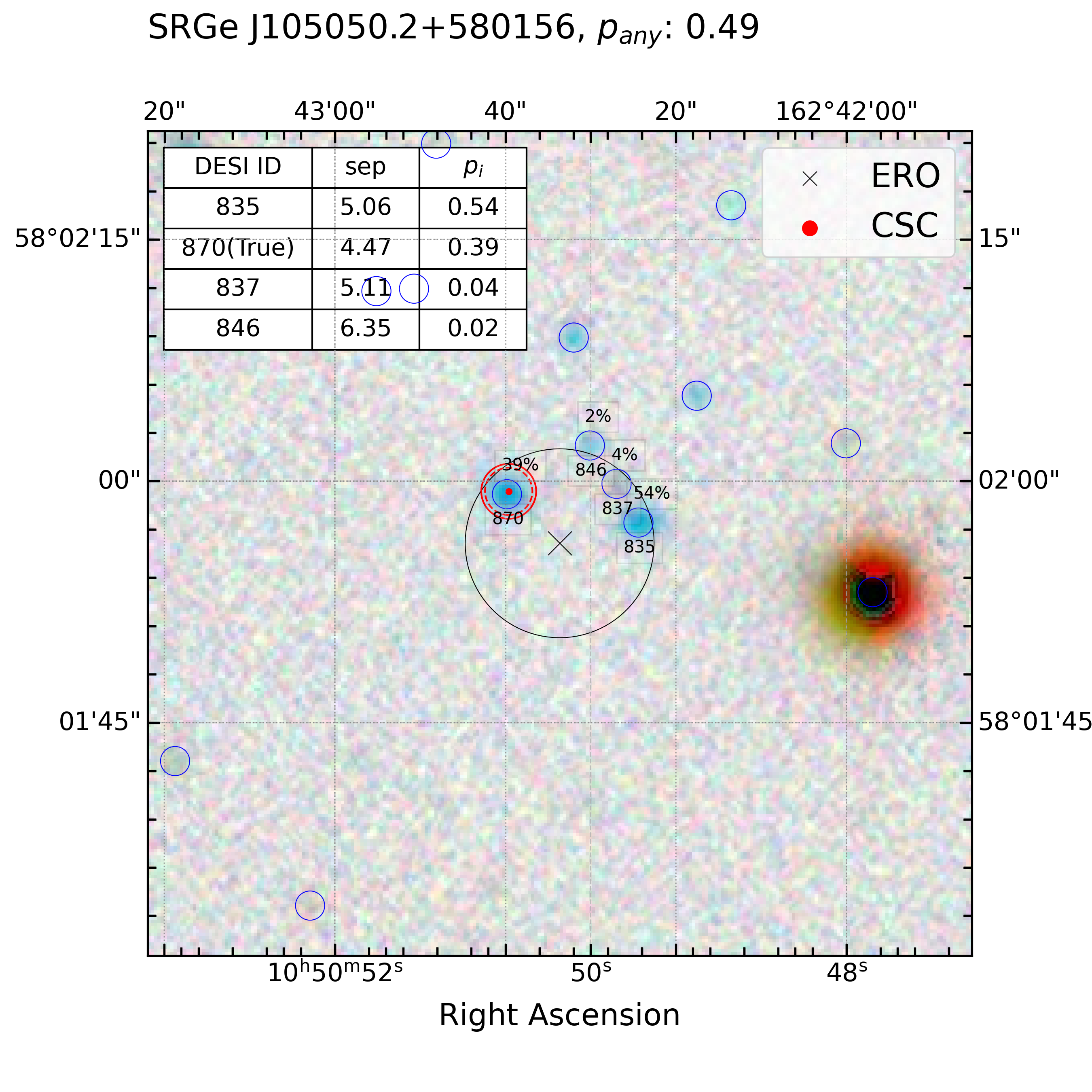}}

\caption{Examples of DESI LIS sky images in the vicinity of eROSITA sources. The cross and the black circle mark eROSITA source position and its positional uncertainty $r_{98}$. The source name and $p_{\rm any}$ are specified in the header. Small blue circles indicate the positions of the DESI LIS objects, the identifier and $p_{\rm i}$ in percent are given below and above each circle, respectively.  Candidates for counterparts with $p_{\rm i}>0.01$, their identifies, and the distance to the eROSITA source in arcsec are specified in tables in the insets. The positions of the Chandra(XMM) sources from  validation catalogue are indicated by red (green) circles. Dotted line indicates radius $r_{\rm false}=1.47"$ around Chandra/XMM. True counterpart is specified by a comment (True). The counterpart identification is correct in cases (a), (b), and (c) and incorrect in case (d), despite relatively large $p_{\rm any}>0.12$.
}
    \label{fig:match_images}
    \end{figure*}

\end{document}